\documentclass{evn2004}
\usepackage[round]{natbib}
\usepackage[scriptsize]{subfigure}
\usepackage{graphicx, amsmath, aas_macros}
\begin{document}
   \title{VLBI detections of a source weaker than $100\,{\rm mJy}$ at 86\,GHz}

   \author{E. Middelberg\thanks{Current address: Australia Telescope National Facility,
           PO Box 76, Epping NSW 1710, Australia},
          \inst{1}
          A. L. Roy,
          \inst{1,2}
          R. C. Walker,
          \inst{3}
          \and
          H. Falcke
          \inst{4}
         }

   \institute{Max-Planck-Institut f\"ur Radioastronomie, Auf dem H\"ugel 69, D-53121 Bonn, Germany\\
              \email{enno.middelberg@csiro.au, aroy@mpifr-bonn.mpg.de}
         \and
              Geod\"atisches Institut der Universit\"at Bonn, Nussallee 17, D-53115 Bonn, Germany
         \and
              National Radio Astronomy Observatory, P.O. Box 0, Socorro, NM, 87801, USA\\
              \email{cwalker@aoc.nrao.edu}
         \and
              ASTRON, P.O. Box 2, 7990 AA Dwingeloo, The Netherlands\\
              \email{falcke@astron.nl}
}

\abstract{We use a new phase-calibration strategy to calibrate the
  phase of 86\,GHz VLBI observations of the FR\,I radio galaxy
  NGC\,4261. Instead of switching between a calibrator source and the
  target source, the target was observed while rapidly switching
  between the target frequency and a lower reference frequency.
  Self-calibration at the reference frequency yielded phase
  corrections which were multiplied with the frequency ratio and
  applied to the target frequency visibilities. The resulting
  detection of NGC 4261 is, to our knowledge, the first of NGC\,4261
  with 86\,GHz VLBI, and it is also the weakest source so far detected
  with VLBI at that frequency.}

\authorrunning{Middelberg et al.}
\titlerunning{VLBI detections of a source weaker than $100\,{\rm mJy}$ at 86\,GHz}

   \maketitle
%

\section{Introduction}

The scientific interest in VLBI observations at 86\,GHz includes
spatially resolved imaging of extragalactic radio jets on the linear
scales on which the jets are launched. The process is not yet
understood and only millimetre VLBI can resolve this region even in
the closest objects. Interests also include investigations of internal
jet structure, jet composition and collimation (e.g.,
\citealt{Doeleman2001}, \citealt{Gomez1999}), and imaging of the event
horizon in, e.g., Sgr\,A$^*$ (\citealt{Falcke2000}).  86\,GHz
observations of the FR\,I radio galaxy NGC\,4261 provide the
opportunity to eventually resolve its jet collimation, because its
combination of proximity and black hole mass yields a linear
resolution of only 200 Schwarzschild radii.

Higher frequency observations involve a number of serious problems.
The sources are usually weaker, the aperture efficiencies of most
radio telescopes used for VLBI drop to 15\,\% or less, the receiver
performances become disproportionately worse, the atmospheric
contribution to $T_{\rm sys}$ becomes larger, and the atmospheric
coherence time scales decrease as $1/\nu$.  In cm VLBI, the coherence
time can be extended using phase-referencing, but in mm VLBI,
phase-referencing is not commonly used, although a successful proof of
concept exists (\citealt{Porcas2002}).  Problems arise from the need
for a suitable, strong phase calibrator in the vicinity of the target
source and relatively long telescope slewing times.

We outlined the technique of fast frequency switching, in
\cite{Middelberg2002}, in which an object is observed while rapidly
switching between the target frequency and a lower reference
frequency. We demonstrated there the feasibility of the technique with
detections of a calibrator and made tentative detections of a weak
target, the AGN in M81.  We have since made considerable improvements
to the observing strategy and present here the first detection of the
AGN in NGC\,4261 with 86\,GHz VLBI.

\section{Principle of phase correction}

The main source of phase noise in VLBI observations at frequencies
higher than about 5\,GHz is turbulence in the troposphere causing
refractive inhomogeneities. The refractive screen is non-dispersive,
and one can self-calibrate the visibility phases at the reference
frequency, $\phi_{\rm r}$, and use the solutions to calibrate the
target frequency visibility phases, $\phi_{\rm t}$, after multiplying
them by the frequency ratio, $r=\nu_{\rm t}/\nu_{\rm r}$. However,
the lag between the two measurements must not exceed half the
atmospheric coherence time.  This is possible with the
VLBA\footnote{The VLBA is an instrument of the National Radio
Astronomy Observatory, a facility of the National Science Foundation,
operated under cooperative agreement by Associated Universities, Inc.} 
because frequency changes need only a few seconds and because the
local oscillator phases return to their original settings after a
frequency switch. After multiplying the phase solutions by the
frequency ratio and applying them to the target-frequency phases,
there remains a constant phase offset between the signal paths at the
two frequencies, $\Delta\Phi$, which must be calibrated.  It can be
monitored with observations of achromatic, strong calibrators, and
must be subtracted from the high-frequency visibility phases. Thus,
the true high-frequency visibilities are phase-referenced to the
source's low-frequency visibilities, and so the technique can prolong
coherence and can measure the position shift of cores in active
galactic nuclei (AGN) with frequency. 

However, in our project, unmodelled ionospheric path length changes,
which are dispersive, limited the coherence to half an hour in the
worst case. Therefore, we here only show how fast frequency switching
can be used to calibrate the short-term phase fluctuations. To remove
the remaining long-term phase drifts and remaining phase offsets, we
used one extra step of self-calibration at the target frequency with a
half-hour solution interval. This extra step of self-calibration
prevented us from making a core-shift measurement.

\section{Observations}

We observed NGC\,4261 as a fast frequency switching target and 3C\,273
and 3C\,279 occasionally for the phase offset, $\Delta\Phi$, and to
test the technique on strong sources.  Dynamic scheduling allowed us
to observe during a period of superb weather, using 256\,Mbps to
record a bandwidth of 64\,MHz with 2-bit sampling.

In this article, a pair of two integrations at $\nu_{\rm r}$ and
$\nu_{\rm t}$ is called a ``cycle'', each integration of which is
called a ``half-cycle'', and a sequence of cycles on the same source
is called a ``scan''.

Several considerations influenced the experiment design. The target
frequency should be an integer multiple of the reference frequency to
avoid having to unwrap phase wraps. We chose a reference frequency of
14.375\,GHz since the third and sixth harmonics at 43.125\,GHz and
86.25\,GHz lie within the VLBA receiver bands. For convenience, we
will refer to these frequencies as ``15\,GHz'', ``43\,GHz'' and
``86\,GHz'', respectively.

We chose a cycle time of 50\,s, of which 22\,s were spent at the
reference frequency of 15\,GHz and the remaining 28\,s were spent at
the target frequency, either 43\,GHz or 86\,GHz.  An average time of
7\,s per half-cycle was lost in moving the subreflector between the
feed horns, resulting in net integration times of 15\,s at $\nu_{\rm
r}$ and 21\,s at $\nu_{\rm t}$. The integration times are a
compromise, depending on source brightness, antenna sensitivity and
expected weather conditions. This setup yielded a $5\,\sigma$
detection limit of 89\,mJy in 15\,s at 15\,GHz for the VLBA on a
single baseline.

\section{Data reduction}

Data reduction was carried out in AIPS. The amplitudes were calibrated
using $T_{\rm sys}$ and gain measurements, and amplitude corrections
for errors in the sampler thresholds were performed using
autocorrelation data. The instrumental delays and inter-IF phase
offsets were corrected using a fringe-finder scan and were found to be
stable over the experiment. A correction for the dispersive delays
introduced by the ionosphere was attempted using GPS-based, global
maps of the total electron content.  Unfortunately, the error in these
maps can be quite high (typically 10\,\% to 20\,\%, but up to 50\,\%),
causing residual phase rates after the transfer of phase solutions
between frequencies. These errors prevented us from making a
core-shift measurement because self-calibration was still required to
calibrate the residual rates.

From the start of a new half-cycle, 5\,s to 8\,s are required to
position the subreflector. Data during that time should be flagged.
Compromise flagging times for each antenna and frequency at the start
of each half-cycle were applied using the AIPS task QUACK.

We fringe-fitted the 15\,GHz data using the AIPS task FRING. We made a
15\,GHz image that we used as a source model in a second run of FRING,
so that the phase solutions did not contain structural phase
contributions.  The solution interval was set to 1\,min, yielding one
phase, delay and phase rate solution per half cycle.  The detection
rate was $\sim90\,\%$. The solution table was written to a text file
to do the phase scaling outside AIPS. A Python program was used to
read in the table, to generate timestamps to coincide with the target
frequency half-cycles, and to calculate a solution consisting of a
phase, a phase rate, and a delay. These solutions were imported to AIPS
to update the most recent calibration table at the target frequency.

\section{Results}

\subsection{43\,GHz}

\begin{figure}[htpb!]
 \includegraphics[width=7cm, angle=270]{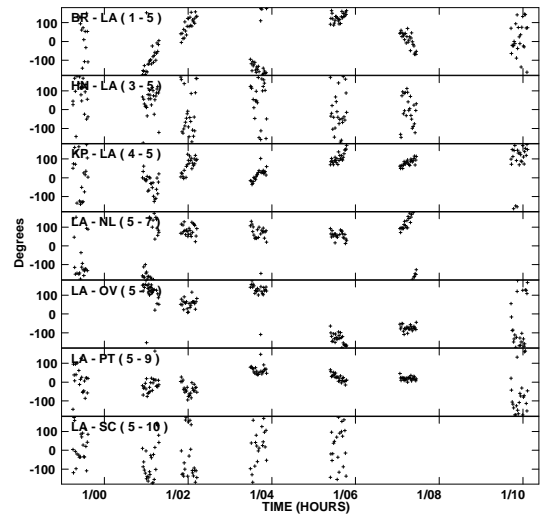}
 \caption{NGC\,4261 calibrated 43\,GHz visibility phases on baselines
 to Los Alamos. Calibration used ionospheric corrections and scaled-up
 phase solutions from fringe-fitting with a clean component model at
 15\,GHz. The weather at HN and SC was worse than at the other
 stations, hence the detections have lower SNR. Each data point is an
 average over a half-cycle.} 
\label{fig:4261_43_tecor+ffs_LA.ps}
\end{figure}

NGC\,4261 was detected on most baselines at all times after scaling
the 15\,GHz solutions to 43\,GHz, with correlated flux densities of
30\,mJy (800\,M$\lambda$) to 160\,mJy (30\,M$\lambda$). Here, the term
``detected'' means that by inspecting the phase-time series by eye one
can see that the phases are not random. The 43\,GHz half-cycle average
visibilities on baselines to Los Alamos are shown in
Fig.~\ref{fig:4261_43_tecor+ffs_LA.ps}. The short-term fluctuations
introduced by the troposphere are almost perfectly calibrated, but
residual phase drifts remain on longer time-scales, especially at the
beginning and at the end of the experiment, when the sun was setting
and rising at Los Alamos and the source elevation was low at most
stations.

\begin{figure}[htpb!]
\centering
 \includegraphics[width=0.9\linewidth, angle=270]{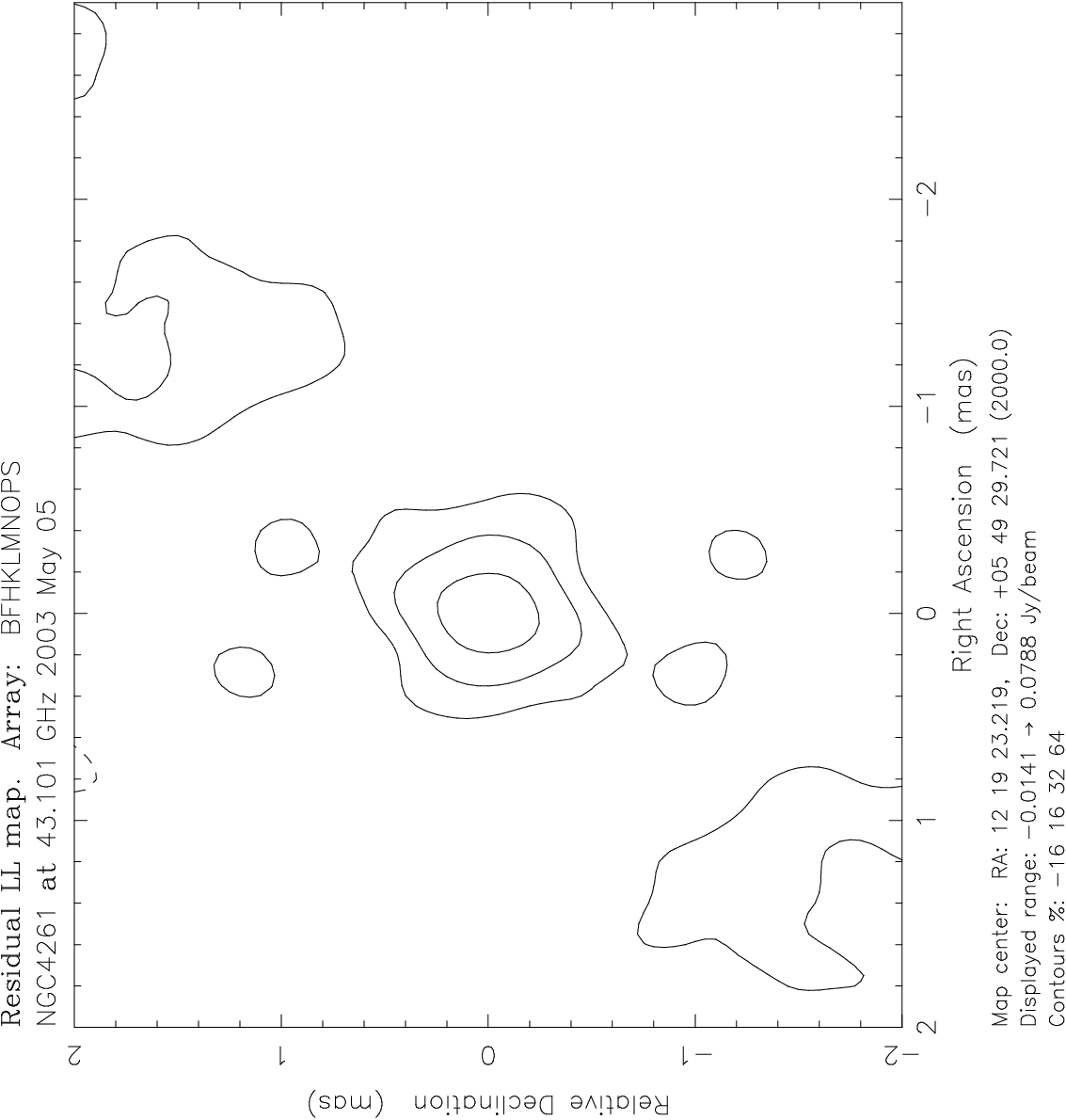}
 \caption[43\,GHz dirty image]{Naturally weighted,
   full-resolution image of \object{NGC\,4261} at 43\,GHz, calibrated
   with scaled-up phase solutions from 15\,GHz. Fringe-fitting has
   been used to solve for one residual phase and rate solution per
   25\,min scan before exporting the data to Difmap. No further
   self-calibration has been applied. The image noise is
   4.4\,mJy\,beam$^{-1}$ and the dynamic range is 18:1. Contours start
   at 12.6\,mJy\,beam$^{-1}$ and increase by factors of 2.}
 \label{fig:NGC4261_43GHz_FFS_raw}
\end{figure}

\begin{figure}[htpb!]
\centering
 \includegraphics[width=0.9\linewidth, angle=270]{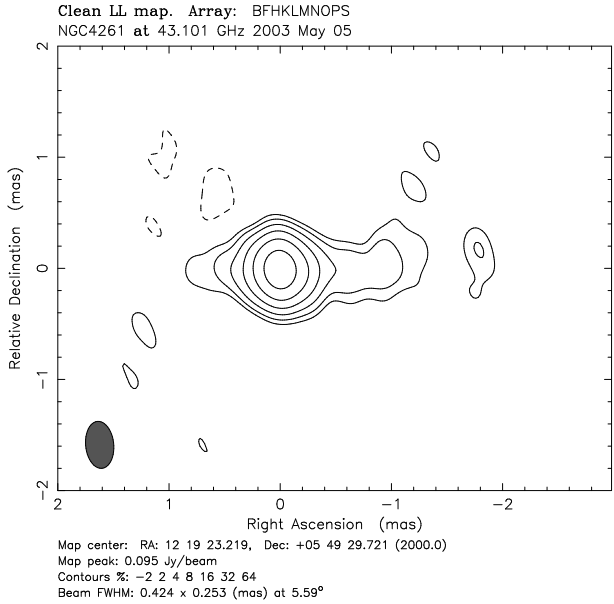}
 \caption[NGC\,4261 43\,GHz clean image]{Data and imaging parameters as
   in Fig.~\ref{fig:NGC4261_43GHz_FFS_raw}, but several cycles of
   phase self-calibration with a solution interval of 30\,s and one
   cycle of amplitude self-calibration with a solution interval of
   12\,h have been applied. The image noise is 0.78\,mJy\,beam$^{-1}$
   and the dynamic range is 122:1. Contours start at
   1.9\,mJy\,beam$^{-1}$ and increase by factors of 2.}
 \label{fig:NGC4261_43GHz_FFS}
\end{figure}

Before making an image from the 43\,GHz visibilities calibrated with
fast frequency switching, the residual phase offsets and phase rates
were removed using fringe-fitting with a solution interval of 30\,min
so that one solution per scan was obtained. The resulting dirty image
(Fig.~\ref{fig:NGC4261_43GHz_FFS_raw}) has a peak flux density of
$79\,{\rm mJy\,beam^{-1}}$ and an rms noise of $4.4\,{\rm
  mJy\,beam^{-1}}$, yielding a dynamic range of 18:1. After several
cycles of phase self-calibration with a solution interval of 30\,s and
one cycle of amplitude self-calibration with a solution interval of
12\,h, the peak flux density was $95\,{\rm mJy\,beam^{-1}}$ and the
rms noise was $0.78\,{\rm mJy\,beam^{-1}}$, so the dynamic range
improved to 122:1 (Fig.~\ref{fig:NGC4261_43GHz_FFS}). The theoretical
rms noise at 43\,GHz was expected to be $0.46\,{\rm mJy\,beam^{-1}}$.
Thus, fast frequency switching allowed us to use a half-hour solution
interval in fringe fitting, and so gain much improved detection
sensitivity.

The performance of the calibration technique can be demonstrated by
comparing the measured phase noise to the expected phase noise. We
have measured the rms of the cycle-to-cycle variations of the
calibrated 43\,GHz visibilities after a) only applying the scaled-up
15\,GHz phase solutions and b) after correcting for the residual phase
errors with a cycle of self-calibration using a 30\,min solution
interval. In case a), the rms was found to be $50^\circ$ in the
beginning and end of the experiment (coinciding with dusk and dawn at
the stations in the south-western US and with predominantly low
elevations), and was $33^\circ$ in the middle of the observations (at
night at most stations and at high elevations). In case b), the rms
dropped slightly to $44^\circ$ and $31^\circ$, respectively.

The expected phase noise in the visibilities calibrated with fast
frequency switching consists of three parts (aside from long-term
changes in electron content, the ionosphere does not contribute
noticeable phase noise, and errors in the source model at $\nu_{\rm
  r}$ are negligible). a) thermal phase noise at the reference
frequency scaled by the frequency ratio, b) thermal phase noise at the
target frequency and c) tropospheric phase changes during the two
integrations.  We estimate the three terms as follows, assuming that
NGC\,4261 has an average compact flux density of 200\,mJy at 15\,GHz
and 100\,mJy at both 43\,GHz and 86\,GHz. a) On a single baseline, the
expected signal-to-noise ratio (SNR) of a detection at 15\,GHz is 11.2
when averaging over the band. In fringe-fitting, this is increased by
$\sqrt{N}$, where $N$ is the number of baselines.  $N\approx8$, so the
SNR of a detection at 15\,GHz increases to 31.7, corresponding to a
phase error of $1.8^{\circ}$.  This error is scaled by the frequency
ratio to $5.4^{\circ}$ at 43\,GHz and $10.9^{\circ}$ at 86\,GHz.  b)
The thermal noise contributions at 43\,GHz and 86\,GHz are
$21.4^{\circ}$ and $54.7^{\circ}$, respectively.  c) We estimated the
tropospheric phase noise within the switching cycle time using
observations of 3C\,273, which are essentially free of thermal noise
and any phase changes during and between the half-cycles are due to
changes in the troposphere. We found the median rms phase noise after
fringe-fitting with a 30\,min solution interval to remove the residual
long-term phase drift to be $13.3^{\circ}$ at 43\,GHz, or
$26.6^{\circ}$ at 86\,GHz.

Adding those three noise components in quadrature yields
$25.8^{\circ}$ at 43\,GHz and is in good agreement with the measured
rms phase noise of $31^{\circ}$.

\subsection{86\,GHz}

\begin{figure}[htpb!]
 \centering
 \includegraphics[width=8cm, angle=270]{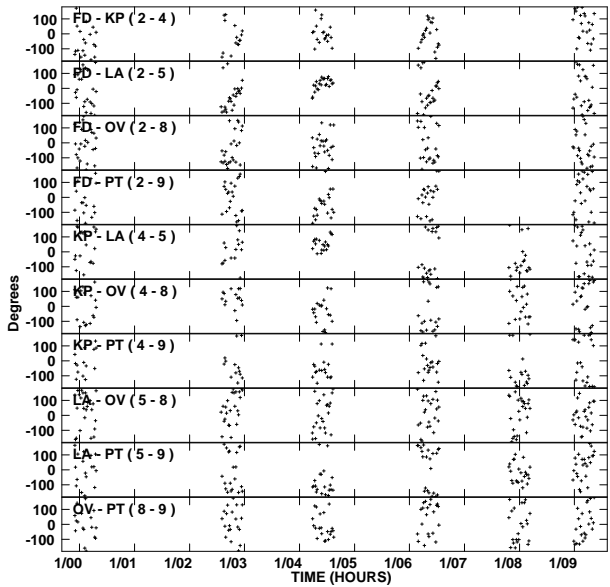}
 \caption{86\,GHz visibility phases on
   baselines among Fort Davis, Kitt Peak, Los Alamos, Owens Valley and
   Pie Town. Calibration has been done with scaled-up phase solutions
   from fringe-fitting at 15\,GHz using a clean component model. Good
   detections were made during almost every 25\,min scan observed at
   night between 2:00\,h UT and 7:00\,h UT.}
 \label{fig:86_GHz_corr_vis}
\end{figure}

\begin{figure}[htpb!]
\centering
 \includegraphics[width=0.9\linewidth, angle=270]{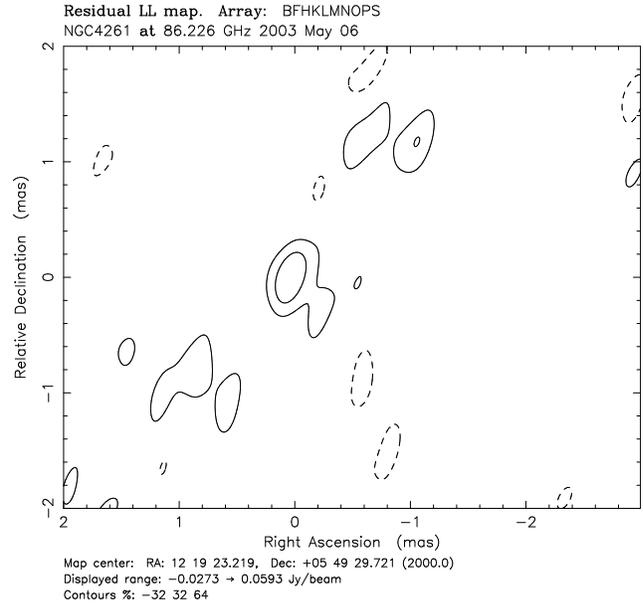}
 \caption{Naturally weighted, full-resolution
   image of \object{NGC\,4261} at 86\,GHz, calibrated with scaled-up
   phase solutions from 15\,GHz. Fringe-fitting has been used to solve
   for one residual phase and rate solution per 25\,min scan before
   exporting the data to Difmap. No further self-calibration has been
   applied. The image noise is 8.4\,mJy\,beam$^{-1}$ and the dynamic
   range is 7:1. Contours are drawn at 19\,mJy\,beam$^{-1}$ and
   38\,mJy\,beam$^{-1}$.}
 \label{fig:4261_86GHz_raw}
 \end{figure}

Following the same data reduction path as for the 43\,GHz data, we
obtained good detections of NGC\,4261 at 86\,GHz on baselines among
the four stations FD, KP, LA, and PT and only weak detections on
baselines to NL, OV and MK. The corrected visibility phases are
plotted in Fig.~\ref{fig:86_GHz_corr_vis}, and an image is shown in
Fig.~\ref{fig:4261_86GHz_raw}.  The peak flux density is 59.3\,mJy. To
our knowledge, this is the first VLBI detection of \object{NGC\,4261}
at 86\,GHz, and is probably the weakest continuum object ever detected
with VLBI at this frequency.

With only delay calibration applied, the median rms phase noise of the
baselines during the best 25\,min scan is $104^{\circ}$, after
applying the scaled 15\,GHz phase solutions is $70^{\circ}$ and after
fringe-fitting with a 30\,min solution interval is $80^{\circ}$ (the
increase in rms phase noise after removal of phase rates has an
unknown cause). We estimate the expected phase noise to
consist of the scaled-up 15\,GHz noise, $10.9^{\circ}$, the thermal
86\,GHz noise, $54.7^{\circ}$, and the tropospheric phase errors,
$26.6^{\circ}$. Their quadrature sum is $61.7^{\circ}$, in agreement
with the measured noise levels.

\section{Summary}

We have demonstrated the feasibility of using fast frequency switching
as a phase calibration method for 86\,GHz VLBI, yielding the detection
of the faintest source so far at that frequency.  Although ionospheric
effects prevented purely phase-referenced images and hence the
detection of a core shift, the technique allows the detection of much
fainter sources at 86\,GHz than is possible with conventional
VLBI. This includes nearby sources which can be imaged with
unprecedented linear resolution and allow one to study the jet
formation and collimation processes.

\bibliography{refs}

\end{document}